\documentclass[%
nofootinbib,
 amsmath,amssymb,
 aps,
 prd,
twocolumn, superscriptaddress
]{revtex4-1}

\usepackage[normalem]{ulem}

\usepackage{graphicx}
\usepackage{dcolumn}
\usepackage{bm}
\usepackage{hyperref}

\usepackage[dvipsnames]{xcolor}
\usepackage{graphicx}
\usepackage{dcolumn}
\usepackage{bm}
\usepackage{orcidlink}
\hypersetup{colorlinks, citecolor=BrickRed}
\hypersetup{colorlinks=true, linkcolor=Blue, urlcolor=MidnightBlue}
\usepackage{hyperref}
\usepackage{ulem}
\usepackage{xcolor}
\usepackage{footnote}

\usepackage{color}

\begin{document}

\preprint{APS/123-QED}

\title{Five-dimensional compact stars in Einstein-Gauss-Bonnet gravity }

\author{Juan M. Z. Pretel}
 \email{juanzarate@cbpf.br}
 \affiliation{
 Centro Brasileiro de Pesquisas F{\'i}sicas, Rua Dr.~Xavier Sigaud, 150 URCA, Rio de Janeiro CEP 22290-180, RJ, Brazil
}

\author{Ayan Banerjee}
 \email{ayanbanerjeemath@gmail.com}
 \affiliation{
 Astrophysics and Cosmology Research Unit, School of Mathematics, Statistics and Computer Science, \\
 University of KwaZulu–Natal, Private Bag X54001, Durban 4000, South Africa 
}

\author{Anirudh Pradhan}
\email{pradhan.anirudh@gmail.com}
\affiliation{Centre for Cosmology, Astrophysics and Space Science, GLA University, Mathura-281 406, Uttar Pradesh, India}

\date{\today}

\begin{abstract}
Within the framework of Einstein-Gauss-Bonnet theory in five-dimensional spacetime ($5D$ EGB), we derive the hydrostatic equilibrium equations and solve them numerically to obtain neutron stars for both isotropic and anisotropic distribution of matter. The mass-radius relations are obtained for SLy equation of state, which describes both the solid crust and the liquid core of neutron stars, and for a wide range of the Gauss-Bonnet coupling parameter $\alpha$. More specifically, we find that the contribution of the Gauss-Bonnet term leads to substantial deviations from Einstein gravity. We also discuss that after a certain value of $\alpha$, the theory admits higher maximum masses compared with general relativity, however, the causality condition is violated in the high-mass region. Finally, our results are compared with the recent observations data on mass-radius diagram.
\end{abstract}

\maketitle


\section{Introduction} 

Einstein's General Relativity (GR) is one of the most successful gravity theories, even after almost one hundred years, that passes successfully with high accuracy all local observational tests both for weak and strong gravity regime \cite{Will2014}. Although successful in describing the observational and the experimental data, there are several unresolved issues which led to an extensive search for alternative gravity theories. In this direction, higher curvature gravity theories are wonderful tools to explore physics beyond the standard model. Their philosophy is based on the metric modification of gravity that generalizes GR in higher dimensions. In particular, Lovelock gravity theories \cite{Lovelock1971, Lovelock1972} are fascinating extensions of GR that include higher curvature interactions while keeping the order of the field equations down to second order in derivatives. Furthermore, such theory is known to be free of ghosts \cite{Zwiebach, Zumino} when expanded on a flat space, and obeys the generalized Bianchi identities which ensure energy conservation.

Our primary objective in this paper is to develop a formalism for a thorough investigation of stellar structure in a broad class of alternative theories to the GR. Here, we will focus our attention on Einstein-Gauss-Bonnet (EGB)  gravity, whose action is given by the Ricci scalar plus the Gauss-Bonnet (GB) term, quadratic in the curvature. This theory is also known to be most general metric torsion-free theory of gravity which leads to conserved equations of motion. In fact, the GB term naturally arises as a low energy effective action of  heterotic string theory  \cite{Wiltshire:1985us, Wheeler:1986}. Interestingly the GB term is a topological invariant in $4D$ spacetime, and hence does not contribute to the gravitational dynamics. Nevertheless, to get a non-trivial contribution, one can generally associate the GB term with a scalar field \cite{Odintsov:2020, Odintsov:2020sqy}.

Since then EGB gravity theories have been studied by many authors over a wide span of years. As a matter of fact, Boulware and Deser \cite{Boulware:1985} first presented spherically symmetric static black hole solution within the framework of EGB gravity. In Refs.~\cite{Cai:2003gr, Cai:2001dz} the thermodynamic properties associated with black hole horizon and  cosmological horizon have been studied for the GB solution in de Sitter and anti-de Sitter (AdS) space. Later on several black hole solutions and their interesting properties have been intensively cultivated by some authors, see e.g.~Refs.~\cite{Ghosh:2016ddh, Rubiera-Garcia:2015yga, Giacomini:2015dwa, Aranguiz:2015voa, Xu:2015eia}. In addition, many fascinating phenomena such as the gravitational collapse of an incoherent spherical dust cloud \cite{Jhingan:2010zz, Maeda:2006pm, Zhou:2011vy, Abbas:2015sja}, geodesic motion of a test particle \cite{Bhawal},  the phase transition of RN-AdS black holes \cite{Xu:2019xif}, Hawking evaporation of AdS black holes \cite{Wu:2021zyl}, radius of photon spheres \cite{Gallo:2015bda}, regular black hole solutions \cite{Ghosh:2018bxg} and wormhole solutions satisfying the energy conditions \cite{Maeda:2008nz, Mehdizadeh:2015jra} have been well studied in the literature. Moreover, some models related to compact objects have been studied in Refs.~\cite{Maharaj:2015gsd, Hansraj:2015tka, Hansraj:2021amz, Hansraj:2019hxh}.

In particular, when $D = 5$ the most general theory leading to second order equations for the metric is the so-called $5D$ EGB theory or Lovelock theory up to second order. The cosmological dynamics of a flat anisotropic multidimensional Universe filled with a barotropic fluid have been investigated by Kirnos and collaborators \cite{Kirnos:2010eng} (see for more \cite{Kirnos:2010eh}). Some authors \cite{Clifton2020} have proposed a constraint on the positive values of GB constant, leading to overall bounds $0 \lesssim \alpha \lesssim 10^2\, \text{km}^{2}$ based on observations of binary black hole systems.  Besides that, many astrophysical solutions have also been found to establish the viability of $5D$ EGB theory, competing different gravity theories. The effects of the GB term on the dynamics of self-gravitating massless scalar spherical collapse has been studied in $5D$ EGB theory \cite{Deppe:2012wk}. In Ref.~\cite{Deppe:2014oua}, the authors have presented the results of numerical simulations of spherically symmetric massless scalar field collapse in $5D$ AdS EGB gravity.  Furthermore, quark stars consisting of a homogeneous and unpaired interacting quark matter were found in Ref.~\cite{Tangphati:2021mvu}. It has been argued that one may achieve quark stars with masses larger than 2$M_{\odot}$ in $5D$ EGB theory for increasing value of coupling constant.

On the other hand, when testing alternative theories one may start from strong-field regime \cite{Psaltis:2008bb}. From this point of view, the formation and evolution of stars can be considered suitable test-beds for higher curvature gravity. Consequently, neutron stars (NSs), one of the most extreme states of matter found in the Universe, are ideal astrophysical environments to constrain gravity theories on strong-field regime. The matter in the inner core of NSs is compressed to densities several times higher than the density of an ordinary atomic nuclei. Nonetheless, the behavior of matter at ultrahigh densities and temperatures in NSs are not fully understood  because such high densities cannot be reproduced in a laboratory conducted on Earth. Only theoretical models and methods can be formulated where there are a very large number of EoS candidates.

In recent measurements of two pulsar masses yielded values close to $2 M_{\odot}$ including the binary millisecond pulsar J1614-2230 \cite{Demorest2010, Fonseca:2016tux} and the pulsar J0348+0432 \cite{Antoniadis2013}, which have provided an important constraint on the EoS at $\rho \gtrsim \rho_{\text{nuc}} $ where $\rho_{\text{nuc}} = 2.8 \times 10^{14}\, \rm g/cm^3$, and tell us about crucial importance of strong interactions in dense matter physics. Furthermore, the mass-radius ($M-r_{\rm sur}$) relations of NSs are very useful because they allow us to understand the complex physical phenomena occurring inside such astrophysical objects. A large enough set of $M-r_{\rm sur}$ measurements is required to determine their influence on other physical properties such as compactness, moment of inertia, spin periods  of rotation, among other astrophysical observables \cite{Lattimer2016, Ozel2016}. However, extensions of GR can have a  substantial impact on the NS macrophysical properties (see e.g. Ref.~\cite{Olmo2020} for a broad review). There is therefore a growing interest not only in restricting the EoS but also in considering viable theories of modified gravity when we study compact stars. It is evident that the EoS and the framework of modified gravity must be constrained from astronomical observations, as for example from multi-messenger observations of the merger GW170817 \cite{Annala2018, Coughlin2018, Radice2018, Creminelli2017, Ezquiaga2017, Baker2017}.

Although it is very common to adopt isotropic perfect fluids to describe the structure of compact stars,  there are strong arguments indicating that the effects of anisotropy cannot be neglected when we deal  with nuclear matter at very high densities and pressures. Within a spherically symmetric context, anisotropic matter means that the interior pressure in the radial direction is different from that in the polar or azimuthal directions. There are some attempts in the literature that suggest the existence of several sources of anisotropy, such as relativistic nuclear interactions \cite{Ruderman, Canuto}, pion condensation \cite{Sawyer:1972cq}, strong magnetic fields \cite{Yazadjiev:2011ks, Cardall:2000bs, Ioka:2003nh}, stellar solid or superfluid cores \cite{Blaschke:2018mqw} or crystallization of the core \cite{Nelmes:2012uf}. Nonetheless, the main causes could be the realization of physics under extreme conditions at the core of NSs, which  was initially pointed out by Bowers and Liang \cite{BowersLiang}.  According to them, for \textit{arbitrary large anisotropy} there is no limiting mass for NS. Theoretical aspects and generalizations of this idea has also been extensively investigated by Ruderman \cite{Ruderman} and showed that nuclear matter tends to become anisotropic at very high densities of order $10^{15}\, \rm g/cm^3$.

It has been shown that a local anisotropic fluid can be considered as an effective single fluid for two-fluid model \cite{HerreraSantos1997} (see also \cite{Harko:2011nu} for a discussion). In Refs.~\cite{Horvat2011, Pretel2020EPJC}, the authors have investigated the impact of anisotropy on NS properties in GR, such as the mass-radius relation and dynamical stability. All static spherically symmetric anisotropic solutions were obtained in Ref.~\cite{Herrera:2007kz}. Moreover, the anisotropic fluid has widely been considered in many astrophysical applications within the context of Einstein gravity, see e.g.~Refs.~\cite{HerreraBarreto, Doneva2012, Isayev2017, Ivanov2017, Maurya2018, Raposo2019, Setiawan:2019ojj, Pretel2019, Rahmansyah2020, Roupas2020, Das2021, Rahmansyah2021} and references therein. The presence of anisotropic pressure may lead to significant changes in the characteristics of relativistic stars even in modified gravity as demonstrated in \cite{Silva2015, Folomeev2018, Mustafa2020, Panotopoulos:2021sbf, Pretel2022CQG}. Concerning this, some authors \cite{Tangphati2021, Jasim2021, Maurya2022EPJP, Maurya2022EPJC} have recently investigated compact stars in the background of $5D$ EGB theory, as well as gravitational collapse \cite{Chatterjee2022}. Therefore, the main goal of the present research is to examine the possibility of using anisotropy to obtain configurations constructed with SLy EoS \cite{DouchinHaensel2001} and to satisfy the current observational data on the $M-r_{\rm sur}$ diagram. Furthermore, we compare our results with the well measured masses of some massive neutron stars reported in the literature such as the millisecond pulsars PSR J1614-2230 \cite{Demorest2010} and PSR J0740+6220 \cite{Cromartie2019}.

The plan of this document is the following: In Sect.~\ref{Sec2} we briefly review EGB gravity in five dimensions and derive the modified TOV equations for stellar equilibrium structure of anisotropic compact stars. Section \ref{Sec3} presents the EoS and two anisotropy models that we use to describe anisotropic NSs. In Sect.~\ref{Sec4} we show our numerical results and analyze the deviations of the physical quantities with respect to standard GR. Finally, in Sect.~\ref{Sec5} we provide our conclusions. We adopt physical units throughout this work.


\section{Modified TOV equations in EGB gravity}\label{Sec2}

Our paper begins with the action of $D$-dimensional Einstein-Gauss-Bonnet (EGB) theory minimally coupled to matter fields, which reads:
\begin{equation}\label{1}
S = \frac{1}{2\kappa}\int d^{D}x\sqrt{-g}\left[ R +\alpha \mathcal{G} \right] + S_m ,
\end{equation}
where $\kappa \equiv 8\pi G/c^4$, $\alpha$ is the GB coupling constant\footnote{This constant has dimensions of length squared and is related to the string tension in string theory.}, $S_m$ stands for the action of standard matter, and the Gauss-Bonnet invariant $\mathcal{G}$ is given by
\begin{equation}\label{2}
\mathcal{G} = R^2 - 4R_{\mu\nu}R^{\mu\nu} +  R_{\mu\nu\sigma\rho}R^{\mu\nu\sigma\rho} .
\end{equation}

We obtain the corresponding field equations by varying the action (\ref{1}) with respect to metric, namely 
\begin{equation}\label{3}
G_{\mu\nu}+\alpha H_{\mu\nu} = \kappa T_{\mu\nu} ,
\end{equation}
with $G_{\mu\nu}$ being the usual Einstein tensor, $H_{\mu\nu}$ the Gauss-Bonnet tensor, and $T_{\mu\nu}$ is the energy-momentum tensor of matter. Such tensor quantities are given by
\begin{eqnarray}
G_{\mu\nu} &=& R_{\mu\nu} - \frac{1}{2}Rg_{\mu\nu} ,  \\
H_{\mu\nu} &=& 2\left[ RR_{\mu\nu} - 2R_{\mu\sigma}R^{\sigma}_{\ \nu} - 2R_{\mu\sigma\nu\rho}R^{\sigma\rho}  \right.  \nonumber  \\
&& \left. - R_{\mu\sigma\rho\lambda}R^{\sigma\rho\lambda}_{\quad \ \ \nu}  \right] - \frac{1}{2}g_{\mu\nu}\mathcal{G} ,  \\
T_{\mu\nu}&=&  -\frac{2}{\sqrt{-g}}\frac{\delta\left(\sqrt{-g}\mathcal{S}_m\right)}{\delta g^{\mu\nu}},
\end{eqnarray}
where $R$ is the Ricci scalar, $R_{\mu\nu}$ the Ricci tensor, and $R_{\mu\nu\sigma\rho}$ is the Riemann tensor. It is easy to see that for $\alpha = 0$, Eq.~(\ref{3}) reduces to the conventional Einstein field equation. Moreover, the GB term does not contribute to the field equations (\ref{3}) for $D \leq 4$.

In order to construct non-rotating neutron stars, the spherically symmetric $D$-dimensional metric describing the interior spacetime is written in the form
\begin{equation}\label{7}
ds^2_D = -e^{2\psi}(cdt)^2 + e^{2\lambda} dr^2 + r^2d\Omega^2_{D-2}, 
\end{equation}
\noindent
where the metric functions $\psi$ and $\lambda$ depend only on the radial coordinate $r$, and $d\Omega^2_{D-2}$ represents the metric on the surface of the $(D-2)$-sphere, namely
\begin{align}\label{8}
    d\Omega^2_{D-2} =&\ d\theta_1^2 + \sin^2\theta_1d\theta_2^2 + \sin^2\theta_1\sin^2\theta_2 d\theta_3^2  \nonumber  \\
    &+ \cdots + \left( \prod_{j=1}^{D-3} \sin^2\theta_j \right)d\theta_{D-2}^2.
\end{align}

In addition, we assume that the matter source is described by an anisotropic fluid, whose energy–momentum tensor is given by
\begin{equation}\label{9}
T_{\mu\nu} = (\epsilon + p_t) u_\mu u_\nu + p_t g_{\mu\nu} - \sigma k_\mu k_\nu ,
\end{equation}
with $u^\mu$ being the $D$-velocity of the fluid, $\epsilon = c^2\rho$ the energy density of standard matter, $\rho$ the mass density, $p_r$ the radial pressure, $p_t$ the tangential pressure, $\sigma \equiv p_t - p_r$ the anisotropy factor, and $k^\mu$ is a unit spacelike $D$-vector along the radial coordinate.

For the line element (\ref{7}) and energy-momentum tensor (\ref{9}), the $00$ and $11$ components of the field equations (\ref{3}) are given by \citep{Tangphati2021}
\begin{align}
    \frac{2}{r}\frac{\lambda'}{e^{2\lambda}} &= \left\lbrace \frac{2\kappa\epsilon}{D-2} -\frac{1- e^{-2\lambda}}{r^2} \left[ (D-3) \right.\right.  \nonumber  \\
    &\left.\left. + \frac{\alpha}{r^2}(D-5)(1 - e^{-2\lambda}) \right]\right\rbrace \left[ 1+ \frac{2\alpha}{r^2}(1- e^{-2\lambda}) \right]^{-1} ,   \label{FieldE1}  \\
    \frac{2}{r}\frac{\psi'}{e^{2\lambda}} &= \left\lbrace \frac{2\kappa p_r}{D-2} + \frac{1- e^{-2\lambda}}{r^2} \left[ (D-3) \right.\right.  \nonumber  \\
    &\left.\left. + \frac{\alpha}{r^2}(D-5)(1 - e^{-2\lambda}) \right]\right\rbrace \left[ 1+ \frac{2\alpha}{r^2}(1- e^{-2\lambda}) \right]^{-1} ,  \label{FieldE2}
\end{align}
where the prime denotes differentiation with respect to the radial coordinate $r$. It is now evident that there is an extra contribution due to higher order terms for $D \geq 5$. Here we assume the particular case $D= 5$, where the quadratic extensions are nontrivial for the stellar structure. Consequently, within the framework of $5D$ EGB gravity, Eq.~(\ref{FieldE1}) leads to an analogous expression as in GR, namely
\begin{equation}\label{12}
\frac{d}{dr}(re^{-2\lambda}) - 1 = -\kappa r^2c^2\rho_{\rm tot} ,
\end{equation}
where $\rho_{\rm tot}$ can be interpreted as a total mass density, defined as
\begin{equation}\label{TotDensiEq}
    \rho_{\rm tot} \equiv \frac{2}{3}\rho  + \frac{e^{-2\lambda}- 1}{\kappa c^2 r^2}\left[ 1+ \frac{4\alpha}{r}\lambda'e^{-2\lambda} \right] .
\end{equation}

To recast these equations into a more familiar form let us now introduce a mass function $m(r)$ through the following relation
\begin{equation}\label{14}
e^{-2\lambda} = 1 - \frac{2Gm}{c^2r} ,
\end{equation}
which can be obtained from the integration of Eq.~(\ref{12}). Thus, the total gravitational mass enclosed by a sphere of radius $r$ can be written as 
\begin{equation}\label{15}
    m(r) = 4\pi\int_0^r \bar{r}^2 \rho_{\rm tot}(\bar{r}) d\bar{r} .
\end{equation}
For a static spherical fluid ball, the energy-momentum conservation equation is given by $\nabla_\mu T^{\mu\nu} =0$, which leads to
\begin{equation}\label{16}
\frac{d\psi}{dr} = -\frac{1}{ c^2\rho + p_r}\frac{dp_r}{dr} + \frac{3 \sigma}{r( c^2\rho + p_r)} .
\end{equation}

Accordingly, by taking into account the field equations (\ref{FieldE1}) and (\ref{FieldE2}) for $D= 5$ together with Eqs.~(\ref{14}) and (\ref{16}), we find that the modified Tolman-Oppenheimer-Volkoff (TOV) equations for $5D$ EGB gravity now read 
\begin{align}
\frac{dm}{dr} =& \left[ \frac{8\pi}{3}r^6\rho - r^3m + \frac{4\alpha G}{c^2}m^2 \right]\left[ r^4 + \frac{4\alpha G}{c^2}rm \right]^{-1} , \label{17}   \\
\frac{dp_r}{dr} =& -2G\left[ \frac{c^2\rho +p_r}{c^2} \right]\left[ \frac{m}{r^2} + \frac{4\pi}{3c^2}rp_r \right]  \nonumber  \\
& \times \left[ 1+ \frac{4\alpha Gm}{c^2r^3} \right]^{-1}\left[ 1 - \frac{2Gm}{c^2r} \right]^{-1} + \frac{3}{r}\sigma .  \label{18}  
\end{align}

The above differential equations are the hydrostatic equilibrium equations for relativistic stars in $5D$ EGB gravity. It is clear that by setting $\alpha$ to zero, these equations reduce to the conventional TOV equations. In order to close the system, an EoS $p_r = p_r(\rho)$ and a specific anisotropy relation for $\sigma$ are required. This enables one to impose the following boundary conditions
\begin{equation}\label{19}
\rho(0) = \rho_c ,  \qquad \qquad  m(0) = 0 ,
\end{equation}
i.e., Eqs.~(\ref{17}) and (\ref{18}) can be integrated for a given central mass density and maintaining the regularity conditions at the center of the star. It is evident that when anisotropy vanishes i.e., $\sigma = 0$, the above system of equations is reduced to the isotropic case.


\section{Equation of state and anisotropy profile}\label{Sec3}

As already mentioned before, to close the system of modified TOV equations and therefore determine the global properties of a NS, it is required to choose an EoS for the micro-physics of dense matter. Although it was initially hypothesized that NSs contain mainly neutrons as if they were a Fermi gas, nowadays we know that the nature and composition of matter in the high-density inner cores of such objects is still unknown. In fact, there are a large number of theoretical models that differ in assumptions about composition, symmetry energy, multi-body interactions, among other inputs \cite{Ozel2016, Guerra2019}. Here we will employ an EoS that involves only nucleonic matter.

In order to construct anisotropic NSs in $5D$ EGB gravity, we follow a procedure analogous to that carried out in Einstein gravity. In other words, one needs to specify a barotropic EoS for radial pressure and also assign an anisotropy function $\sigma \equiv p_t - p_r$ since there is an extra degree of freedom. Nonetheless, it is worth noting that some authors have adopted an EoS for both radial and tangential pressure, see e.g. Refs.~\cite{Abellan2020a, Abellan2020b} for discussions on the matter.

For the micro-physical relation between mass density and radial pressure of stellar fluid, we use the well-known SLy EoS \cite{DouchinHaensel2001} which is compatible with the constraints extracted from the event GW170817 --- the first detection of gravitational waves from a binary neutron star inspiral \cite{Abbott}. This unified EoS describes both the thin crust and the massive liquid core of NSs, and it can be represented by the following equation \citep{HaenselPotekhin2004}
\begin{eqnarray}\label{20}
\zeta(\xi) &=& \frac{a_1 + a_2\xi + a_3\xi^3}{1 + a_4\xi}f(a_5(\xi - a_6))  \nonumber  \\
&& + (a_7 + a_8\xi)f(a_9(a_{10} - \xi))   \nonumber  \\
&& + (a_{11} + a_{12}\xi)f(a_{13}(a_{14} - \xi))  \nonumber  \\
&& + (a_{15} + a_{16}\xi)f(a_{17}(a_{18} - \xi)) ,
\end{eqnarray}
where $\zeta \equiv \log(p_r/ \rm dyn\ cm^{-2})$, $\xi \equiv \log(\rho/ \rm g\ cm^{-3})$ and $f(x) \equiv 1/(e^x +1)$. The values $a_i$ are fitting parameters and can be found in \citep{HaenselPotekhin2004}. Besides this EoS, we will employ two anisotropy profiles for $\sigma$ provided in the literature to model anisotropic matter at very high densities and pressures. In what follows we will describe in more detail the functional relations for the anisotropy.

The first anisotropy profile adopted in this work is the model proposed by Horvat \textit{et al.} in Ref.~\cite{Horvat2011}, given by
\begin{equation}\label{HorvatEq}
    \sigma = \beta_{\rm H}p_r\mu = \beta_{\rm H}p_r (1 - e^{-2\lambda}) ,
\end{equation}
where $\beta_{\rm H}$ is a dimensionless parameter that measures the amount of anisotropy inside the star and, in principle, it can assume positive or negative values of the order of unity \cite{Doneva2012, Folomeev2018, Silva2015, Yagi2015, Pretel2020EPJC}. Such anisotropy model is also known as quasi-local ansatz because the compactness $\mu$ denotes a quasi-local variable. Since $\mu \rightarrow 0$ when $r \rightarrow 0$, the relation (\ref{HorvatEq}) guarantees that the anisotropy factor vanishes at the stellar center, i.e.~the fluid becomes isotropic. Furthermore, in the Newtonian limit (when the pressure contribution to the energy density is negligible) the effect of anisotropy vanishes in the hydrostatic equilibrium equation.

Following Bowers and Liang \cite{BowersLiang}, we will consider another form of the anisotropy factor, 
\begin{equation}\label{Bowers-Liang}
    \sigma = \frac{\beta_{\rm BL}G}{c^4}(c^2\rho + p_r)(c^2\rho + 3p_r)e^{2\lambda}r^2 ,
\end{equation}
for a non-rotating configuration. Here, $\beta_{\rm BL}$ is a constant parameter that quantifies the amount of anisotropy, and we recover the isotropic case when $\beta_{\rm BL} = 0$. Moreover, in the non-relativistic regime the effect of anisotropy does not vanish in the hydrostatic equilibrium equation, which could be an unphysical trait as argued in Ref.~\cite{Yagi2015}. The anisotropy parameter $\beta_{\rm BL}$ assumes values similar to $\beta_{\rm H}$ \cite{Folomeev2018, Silva2015, Yagi2015, Biswas2019, Pretel2020EPJC}.

\begin{table}
\caption{\label{table1} 
Isotropic NSs with central mass density $\rho_c = 2.0 \times 10^{18}\, \rm{kg}/\rm{m}^3$ corresponding SLy EoS in $5D$ EGB gravity for different values of the coupling constant. The corresponding graphs for mass function and pressure are plotted in Fig. \ref{figure1}. The difference in mass between the values obtained in GR (in $4D$) and $5D$ EGB gravity is given by $\Delta M$. }
\begin{ruledtabular}
\begin{tabular}{cccc}
 Theory: $\alpha$ [$\rm km^2$]  &  $r_{\rm sur}$  [\rm{km}]  &  $M$ [$M_\odot$]  &  $\Delta M$ [$M_\odot$]  \\
\colrule
$-10$  &  12.827  &  1.239  &  0.751  \\
$-5$  &  13.044  &  1.330  &  0.660  \\
$0$  &  13.256  &  1.418  &  0.572  \\
$5$  &  13.463  &  1.504  &  0.486  \\
$10$  &  13.664  &  1.587  &  0.403  \\
$20$  &  14.047  &  1.746  &  0.244  \\
\end{tabular}
\end{ruledtabular}
\end{table}

\begin{figure*}
 \includegraphics[width=8.455cm]{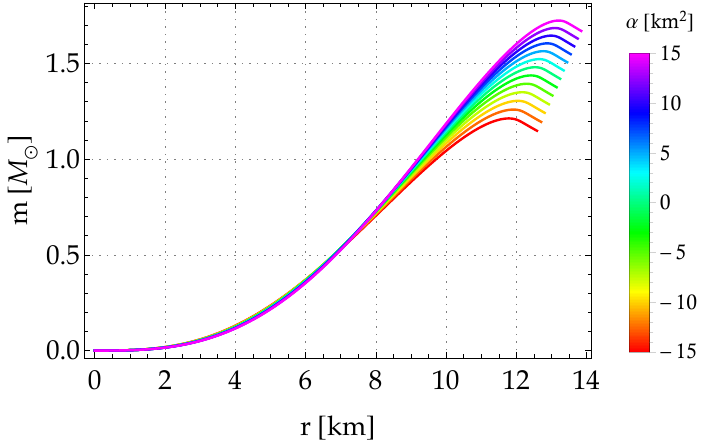} 
 \includegraphics[width=8.2cm]{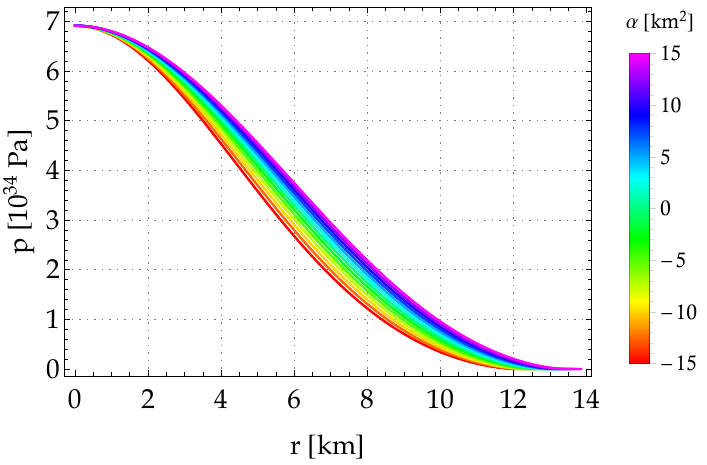}
 \caption{\label{figure1} Mass parameter (left panel) and pressure of the isotropic fluid (right panel) as functions of the radial coordinate within the framework of $5D$ EGB theory of gravity. These plots correspond to a central mass density $\rho_c = 2.0 \times 10^{18}\, \rm kg/m^3$ with SLy EoS (\ref{20}) for $\sigma =0$ and different values of the coupling constant $\alpha$. The gravitational mass (at the surface) of the star increases with increasing value of $\alpha$. Interestingly, the mass suffers a small drop after reaching a maximum near the surface for any value of $\alpha$. For some specific calculations of surface radius and mass, see also Table \ref{table1}.}
\end{figure*}

\begin{figure*}
 \includegraphics[width=8.336cm]{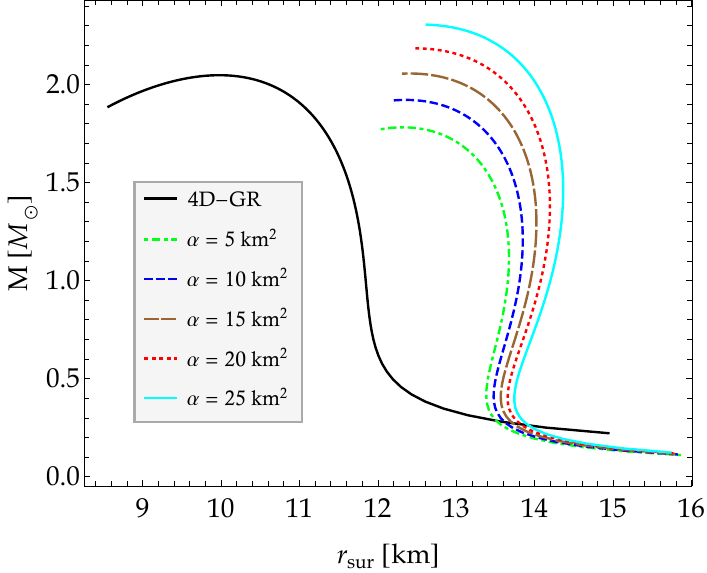} 
 \includegraphics[width=8.2cm]{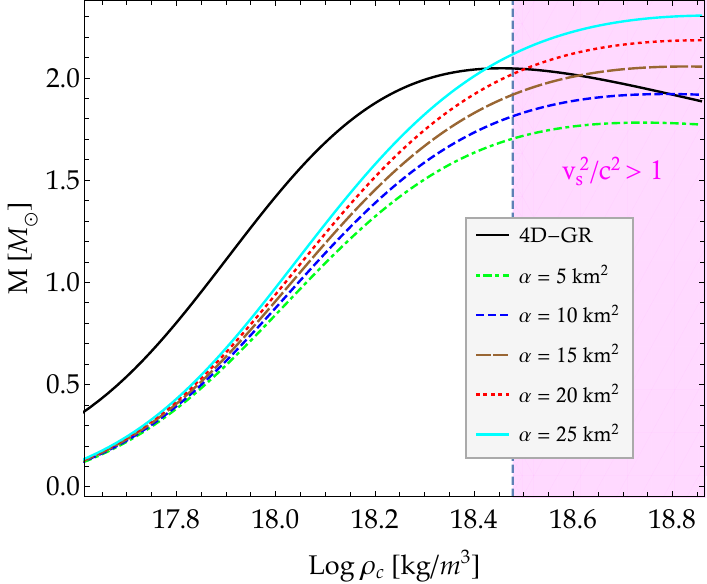}
 \caption{\label{figure2} Left panel: Mass-radius diagram in $5D$ EGB gravity for several values of $\alpha$, where $M$ represents the total gravitational mass on the surface of the star. Right panel: Mass-central density relation, where the magenta region indicates that the sound speed is greater than the speed of light in neutron-star matter with SLy EoS. The standard GR solution is shown in both plots as a benchmark by a solid black line. Note that the solutions corresponding to the EGB gravity generate maximum-mass configurations that do not satisfy the causality condition at the stellar center. }
\end{figure*}


\section{Numerical results and discussion}\label{Sec4}

\subsection{Isotropic configurations}

Let us begin by examining the isotropic case (i.e., when $p_t = p_r = p$) within the context of $5D$ EGB gravity. To do so, we numerically integrate the modified TOV equations (\ref{17}) and (\ref{18}) with boundary conditions (\ref{19}) from the center at $r=0$ up to the stellar surface at $r = r_{\rm sur}$ where the pressure vanishes. In addition, we set $\sigma =0$ and specify a value for the coupling constant $\alpha$. In particular, for a central mass density $\rho_c = 2.0 \times 10^{18}\, \rm kg/m^3$ with SLy EoS (\ref{20}), we obtain the solutions shown in Fig.~\ref{figure1}. This figure shows typical values for the star's mass $M \sim (1.2 - 1.7)\, M_{\odot}$ corresponding to their isotropic pressure. As one can see the properties of such stars change considerably when we vary the  parameter $\alpha$. Indeed, as $\alpha$ increases the surface radius and total mass of the star also increases. In Table \ref{table1}, we show how much the mass obtained in EGB gravity varies with respect to its corresponding GR value. It is noticeable that for a given value of $\alpha$ in the left plot of Fig.~\ref{figure1}, the mass function suffers a small drop after reaching a maximum value near the surface. This phenomenon is associated with only the EGB gravity, not with its standard GR counterpart (where the mass function is always increasing as we approach the surface of the star).

By solving the stellar structure equations for a range of central density values, we can get a family of NSs in $5D$ EGB gravity. Figure \ref{figure2} illustrates the mass-radius and mass-central density relations for the SLy nucleonic EoS and  different values of $\alpha$. We also showed that for the GR case a smaller central density yields a less massive NS and would lead to less compact object. In this perspective, the GB term plays an important role for EGB theory of gravity identifying a substantial deviation on $M-r_{\rm sur}$ relations at which EGB results differ significantly from standard GR in $4D$. As a peculiar feature of EGB gravity is that the greatest effect of the parameter $\alpha$ on the global properties of a NS takes place in the high-mass region, while the variations of $\alpha$ produce irrelevant changes for small masses (i.e., at low central energy densities). Here it is important to highlight that this qualitative behavior is similar for isotropic non-magnetized NSs in four-dimensional EGB gravity \cite{Bordbar2024}.

One observes that the maximum-mass points on the mass versus central density curves can even exceed the GR counterpart from a certain value of the coupling constant. However, it has been argued that any stellar model must satisfy general physical requirements or ``physical acceptability conditions'' \cite{Mak2003, Abreu2007}, so that causality violations are typically seen as unrealistic and undesirable features of the adopted model. Indeed, the strictly relativistic position is that compact star models which violate the causality requirement $v_s^2 \leq c^2$ are ruled out as unphysical \cite{Ellis2007}. According to the right panel of Fig.~\ref{figure2}, we see that there is a region where the speed of sound (defined by $v_s \equiv \sqrt{dp/d\rho}$) is greater than the speed of light $c$. This violates the causality condition, and thus it is not possible to describe physically realistic massive NSs in EGB gravity considering only isotropic pressures. Perhaps the superluminal signal propagation inside massive NSs in $5D$ EGB gravity could be avoided by considering other EoSs for dense matter.

On the other hand, the stellar configurations presented in Fig.~\ref{figure2} describe NSs in hydrostatic equilibrium, however, such equilibrium can be stable or unstable with respect to a small radial perturbation. It has been shown, at least in GR (see e.g.~Refs.~\cite{Glendenning, Haensel2007}), that a turning point from stability to instability occurs when $dM(\rho_c)/d\rho_c = 0$. This means that the stable branch in the sequence of stars is located before the critical density corresponding to the maximum-mass point. Consequently, the stable stars in the right panel of Fig.~\ref{figure2} are found in the region where $dM(\rho_c)/d\rho_c >0$. Due to its simplicity, this condition has been widely used in the literature. Nevertheless, we must point out that such a condition is just necessary but not sufficient to determine the limits of stellar stability. 

It is well known in GR that the existence of anisotropic pressure leads to more massive compact stars. Thus in the next section we are going to explore the consequences of including anisotropy in the stellar structure within the framework of $5D$ EGB gravity.

\begin{table}
\caption{\label{table2} 
Measurements of masses for the most massive neutron stars observed in nature.}
\begin{ruledtabular}
\begin{tabular}{cc}
Source and reference   &   Measured mass [$M_\odot$]  \\
\colrule
PSR J1614-2230 \cite{Demorest2010}  &  $ 1.97\pm 0.04 $  \\
PSR J0348+0432 \cite{Antoniadis2013}  &  $ 2.01\pm 0.04 $  \\
PSR J0740+6620 \cite{Cromartie2019}  &  $ 2.14_{-0.09}^{+0.10} $  \\
PSR 2215+5135 \cite{Linares2018}  &  $ 2.27_{-0.15}^{+0.17} $  \\
\end{tabular}
\end{ruledtabular}
\end{table}

\begin{figure*}
 \includegraphics[width=7.8cm]{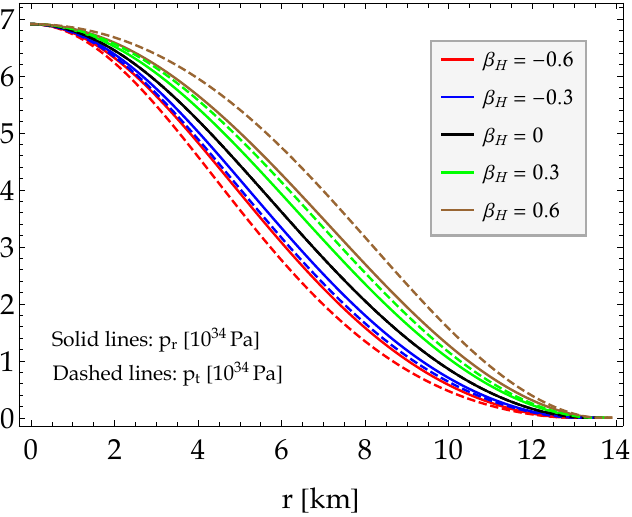} \ \
 \includegraphics[width=7.75cm]{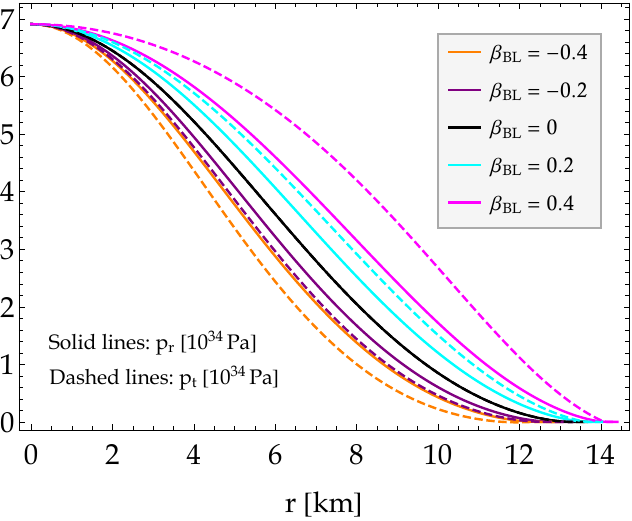}
 \caption{\label{figure3} Numerical solution of the system of equations (\ref{17}) and (\ref{18}) for a NS with central density $\rho_c = 2.0 \times 10^{18}\, \rm kg/m^3$ and  SLy EoS in $5D$ EGB gravity with $\alpha = 10\, \rm km^2$ for both plots. We used the quasi-local ansatz (left panel) and Bowers-Liang profile (right panel) for different values of the anisotropy parameters $\beta_{\rm H}$ and $\beta_{\rm BL}$. The solid and dashed curves correspond to the radial and tangential pressures, respectively. For each profile, larger values of $\vert \beta\vert$ correlate with higher amounts of anisotropy in the intermediate regions of the star. We also note that the anisotropy vanishes both at the center and at the surface of the star, as expected. }
\end{figure*}

\begin{figure*}
 \includegraphics[width=8.2cm]{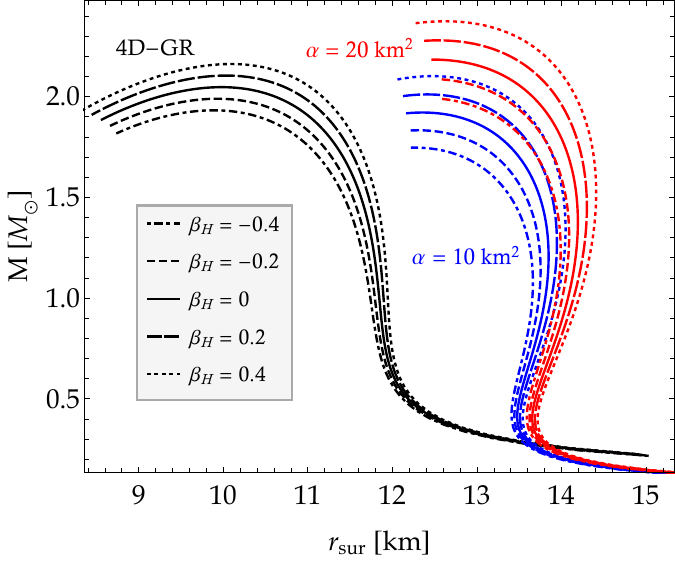} \
 \includegraphics[width=8.24cm]{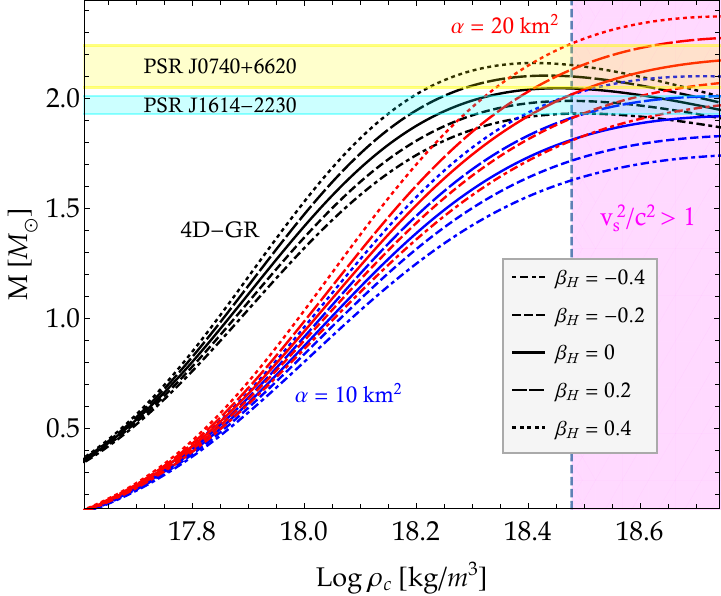}
 \caption{\label{figure4} Illustration of the macro-physical properties predicted by the $5D$ EGB gravity. Mass versus radius curves (left panel) and mass versus central density relations (right panel) for anisotropic NSs with SLy EoS for radial pressure and anisotropy profile (\ref{HorvatEq}). The standard GR solution has been included for comparison reasons. As we can see, both radius and mass undergo significant changes due to the influence of the GB term as well as the anisotropy parameters $\beta_{\rm H}$.  The horizontal bands in cyan and yellow colors stand for the observational measurements of the NS masses reported in Refs. \cite{Demorest2010} and \cite{Cromartie2019}, respectively. The magenta region indicates that $v_s^2 \equiv dp_r/d\rho > c^2$ at the stellar center.  }
\end{figure*}

\begin{figure*}
 \includegraphics[width=8.2cm]{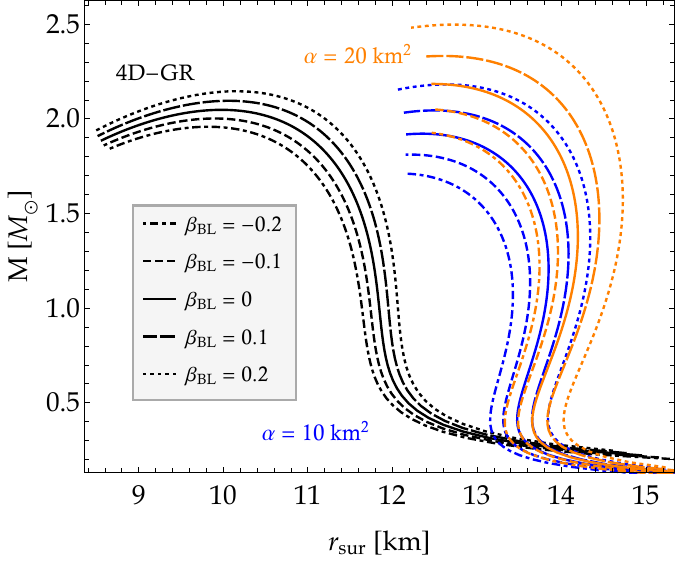} \
 \includegraphics[width=8.24cm]{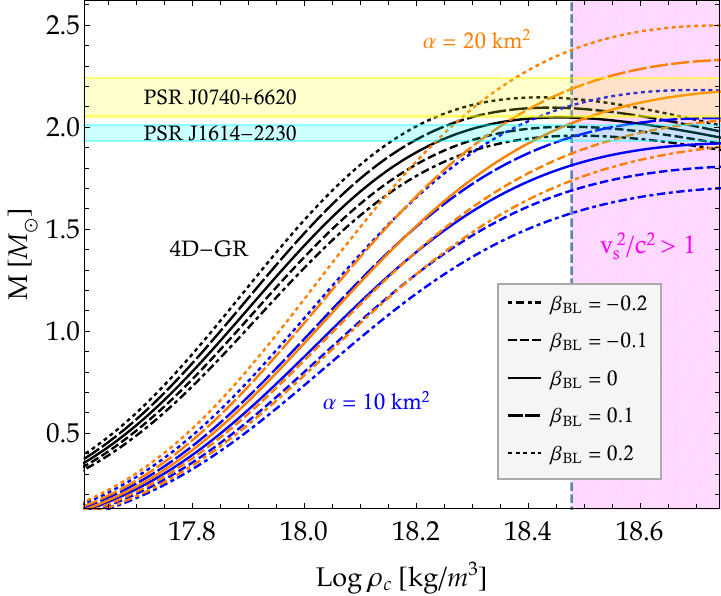}
 \caption{\label{figure5} Mass-radius diagrams (left panel) and mass-central density relations (right panel) for anisotropic neutron stars in $5D$ EGB gravity with anisotropy profile (\ref{Bowers-Liang}). Note that the behavior in the curves is qualitatively similar to the results of Fig. \ref{figure4}, although the values of $\beta_{\rm BL}$ are smaller than $\beta_{\rm H}$. Furthermore, at low central densities, the quasi-local ansatz introduces greater changes in the gravitational mass than the Bowers-Liang ansatz. The colored regions represent the same as in Fig. \ref{figure4}. }
\end{figure*}


\subsection{Anisotropic configurations}

Bearing in mind that our aim is also to explore the effect of anisotropic pressure on NSs, here we include an anisotropy factor $\sigma$ in the TOV equations. In other words, we numerically solve the system of Eqs.~(\ref{17}) and (\ref{18}) by setting the boundary conditions (\ref{19}), and specifying the values of coupling constant $\alpha$ as well as the anisotropy parameter $\beta$. We use the anisotropy profiles (\ref{HorvatEq}) with $\beta_{\rm H}$ and (\ref{Bowers-Liang}) with $\beta_{\rm BL}$, where the tangential pressure $p_t$ differs from the radial one $p_r$ through an ansatz. For each model under consideration, the isotropic case can be recovered when $\beta \rightarrow 0$. 

The radial profile of each model is illustrated in the plots of Fig.~\ref{figure3} for a central density $\rho_c = 2.0 \times 10^{18}\, \rm kg/m^3$, $\alpha = 10\, \rm km^2$, and for several values of $\beta$. Each color indicates a specific value of the anisotropy parameter, and solid and dashed lines stand for radial and tangential pressure, respectively. It is interesting to observe that for the two models adopted here the anisotropy vanishes at the center (which is a required condition in order to guarantee regularity), is highest in the intermediate regions, and it vanishes again at the stellar surface. These results revel that anisotropies have a similar qualitative behaviour for both ansatze.

For the anisotropy function (\ref{HorvatEq}), Fig.~\ref{figure4} displays the mass-radius and mass-central density relations for anisotropic NSs with SLy EoS in EGB gravity for two particular values of the coupling constant $\alpha$. It is evident that for lower values of coupling constant the maximum masses and their corresponding maximum radius have lower values, which is similar to the isotropic case. Nonetheless, the anisotropy parameter $\beta_{\rm H}$ introduces relevant changes in both mass and radius, mainly in the high-central-density region. Positive (negative) values of $\beta_{\rm H}$ generate higher (lower) maximum masses with respect to the isotropic case. This means that the higher-order curvature terms (i.e., the quadratic GB term) and the influence of anisotropies give rise to massive NSs whose results are in good agreement with the observational constraints on millisecond pulsar that coming from different astrophysical sources. Finally, in Table \ref{table2}, we present the numerical values for the systems under consideration and show that our results are consistent with the recent observational data. However, once again we have to point out that the maximum-mass configurations are found in the region where the causality condition is violated.

Finally, in Fig.~\ref{figure5}, we demonstrate the mass-radius and mass-central density diagrams for the anisotropy profile (\ref{Bowers-Liang}). It can be noted that the behavior is qualitatively similar to the results generated by ansatz (\ref{HorvatEq}), although the values of $\beta_{\rm BL}$ are smaller than $\beta_{\rm H}$. As a result, an action containing higher-order curvature terms (with suitable values of the parameter $\alpha$) together with the presence of anisotropy allow us to obtain maximum masses greater than $2.0\, M_\odot$.


\section{Conclusions}\label{Sec5}

In this work we have studied static neutron stars within the context of 5-dimensional Einstein-Gauss-Bonnet theory of gravity. The GB term (built out of quadratic contractions of the Riemann and Ricci tensors) generates a nontrivial extra contribution that ends up influencing the internal structure of the stars and, consequently, modifies the mass-radius relations. Initially we have obtained numerical solutions corresponding to stellar configurations described by an isotropic fluid where the degree of modification with respect to Einstein gravity is measured by the GB coupling constant $\alpha$. Particularly, under a certain value of $\alpha$, the radius increases and the total gravitational mass decreases. Furthermore, it is possible to obtain larger maximum masses for certain positives values of $\alpha$, however, such configurations violate the causality condition.

In addition, we have explored the role of anisotropies in the stellar structure by introducing an extra degree of freedom $\sigma$ in which the free parameter $\beta$ is set with the help of some phenomenological ansatz. The parameter $\beta$ measures the deviation from the isotropic fluid. We analyzed how the GB term and the fluid anisotropies modify the two most basic properties of NSs. We have shown that the effect of positive anisotropic pressure is mainly to increase the mass of these stars, while negative anisotropies have an opposite impact. This finding confirms that NSs are sensitive to the choice of anisotropic model even in $5D$ EGB gravity. Of course, as in standard Einstein gravity, we should remark that this depends on the amount of anisotropy inside the star. Furthermore, it is important to emphasize that this article is far from being a complete picture of the problem in $D>4$ and perhaps the term ``neutron stars'' should be extended to five or more dimensions. We have invoked the specific SLy EoS in order to be able to close the system of differential equations, and it may also be necessary to approach the microphysics of the star from a 5-dimensional point of view, such as the generalized version of the MIT bag model EoS in $d$-dimensions \cite{Arbanil2019}. We will leave this discussion for a future work, although our study already opens several new aspects in the future analysis of compact stars in higher dimensions within the framework of EGB gravity.

Since the standard GR does not reveal the existence of super-massive NSs using a soft EoS (for instance, the SLy EoS favored by GW170817), it becomes interesting to explore extensions of Einstein gravity. Concerning this, we have studied how the values of $\alpha$ and $\beta$ would affect the gravitational mass of a NS, namely, how the Gauss-Bonnet Lagrangian and anisotropies would give rise to smaller and larger masses with respect to the conventional GR results. We expect that future astronomical observations will allow us to impose tight constraints on the coupling constant $\alpha$ in $5D$ EGB gravity. Using our inference of the maximum NS mass, one may be able to identify what is the degree of modification with respect to GR. It is pertinent to mention that the results corresponding to pure Einstein gravity have been included for comparison reasons, which help us to get a better idea of the new findings generated by the $5D$ EGB term on the mass-radius diagram.

\begin{acknowledgments}
JMZP acknowledges support from ``Fundação Carlos Chagas Filho de Amparo à Pesquisa do Estado do Rio de Janeiro'' -- FAPERJ, Process SEI-260003/000308/2024.
\end{acknowledgments}\


%

\end{document}